\documentclass[aps,prl,superscriptaddress,twocolumn,showpacs,amsmath,amssymb,letter]{revtex4}

\usepackage{graphicx}
\usepackage{dcolumn}
\usepackage{bm}
\usepackage[makeroom]{cancel}

\begin{document}

\title{Statistical mechanics guides the motions of cm scale objects}

\author{S. Siriroj}
\affiliation {School of Physics, Suranaree University of Technology, Nakhon
Ratchasima, 30000, Thailand}
\author{P. Simakachorn}
\affiliation {School of Physics, Suranaree University of Technology, Nakhon
Ratchasima, 30000, Thailand} \affiliation{SCiUS program, Suranaree University
of Technology and Rajsima Wittayalai School, Nakhon Ratchasima, 30000,
Thailand}
\author{N. Khumtong}
\affiliation{SCiUS program, Suranaree University of Technology and Rajsima
Wittayalai School, Nakhon Ratchasima, 30000, Thailand}
\author{T. Sukhonthamethirat}
\affiliation{SCiUS program, Suranaree University of Technology and Rajsima
Wittayalai School, Nakhon Ratchasima, 30000, Thailand}
\author{S. Chaiyachad}
\author{P. Chanprakhon}
\author{K. Chanthorn}
\author{S. Dawprateep}
\author{T. Eknapakul}
\author{I. Fongkaew}
\author{C. Jaisuk}
\author{T. Jampreecha}
\author{W. Jindata}
\author{Y. Kaeokhamchan}
\author{T. Kongnok}
\author{P. Laohana}
\author{K. Lapawer}
\author{S. Lowpa}
\author{A. Mooltang}
\author{S. Musikajaroen}
\author{S. Nathabumroong}
\author{A. Panpar}
\author{S. Phumying}
\author{S. Polin}
\author{A. Rasritat}
\author{A. Ritwiset}
\author{W. Saengsui}
\author{W. Saenrang}
\author{T. Saisopa}
\author{S. Sangphet}
\author{T. Sawasdee}
\author{S. Sonsupap}
\author{S. Suksombat}
\author{T. Suyuporn}
\author{M. Tepakidareekul}
\author{T. Thiwatwaranikul}
\author{P. Tipsawat}
\author{S. Waiprasoet}
\author{W. Meevasana}
\email[Authors to whom correspondence should be addressed. Electronic
addresses: ]{worawat@g.sut.ac.th} \affiliation {School of Physics, Suranaree
University of Technology, Nakhon Ratchasima, 30000, Thailand} \affiliation
{Thailand Center of Excellence in Physics, CHE, Bangkok 10400, Thailand}

\date{\today}

\begin{abstract}
Calculations and mechanistic explanations for the probabilistic movement of
objects at the highly relevant cm to m length scales has been lacking and
overlooked due to the complexity of current techniques. Predicting the
final-configuration probability of flipping cars for example remains
extremely challenging. In this paper we introduce new statistical
methodologies to solve  these challenging macroscopic problems. Boltzmann's
principles in statistical mechanics have been well recognized for a century
for their usefulness in explaining thermodynamic properties of matter in
gas, liquid and solid phases. Studied systems usually involve a large
number of particles (e.g. on the order of Avogadro's number) at the atomic
and nanometer length scales. However, it is unusual for Boltzmann's
principles to be applied to individual objects at centimeter to human-size
length scales. In this manuscript, we show that the concept of statistical
mechanics still holds for describing the probability of a tossed
orthorhombic dice landing on a particular face. For regular dice, the one
in six probability that the dice land on each face is well known and easily
calculated due to the 6-fold symmetry. In the case of orthorhombic  dice,
this symmetry is broken and hence we need new tools to predict the
probability of landing on each face. Instead of using classical mechanics
to calculate the probabilities, which requires tedious computations over a
large number of conditions, we propose a new method based on Boltzmann's
principles which uses synthetic temperature term. Surprisingly, this
approach requires only the dimensions of the thrown object for calculating
potential energy as the input, with no other fitting parameters needed. The
statistical predictions for landing fit well to experimental data of over
fifty-thousand samplings of dice in 23 different dimensions. We believe
that the ability to predict, in a simple and tractable manner, the outcomes
of macroscopic movement of large scale probabilistic phenomena opens up a
new line of approach for explaining many phenomena in the critical
centimeter-to-human length scale.
\end{abstract}

\maketitle

Tossing a die from a significant height on a hard floor is generally thought
as a random process since its motion is unpredictable from the initial state
of the die. However, regardless of its random motion, one can accurately
predict that the probability of a die landing on any particular face of a
standard cubic die is 1/6. This predictability is the result of the six-fold
symmetry of the die alone. It is non-trivial to ask, "how will this 1/6
probability change if we break this six-fold symmetry?" The motion of a
macroscopic object is in principle governed by classical mechanics and
Newton’s laws. If all the necessary initial conditions are given, the
equations of motion can be written for calculating the coordinates and
momenta of an object anytime thereafter. In practice, the bouncing motion of
simple objects, e.g. dice and coins, is already difficult to calculate. We
are not aware of any previous work studying the motion of a bouncing die
using the full die geometry, although the bouncing motion of a simplified
object such as a barbell has already been reported by Nagler and Richter
\cite{barbell}. The calculated phase space of the final configuration of the
barbell (i.e. whether the barbell points to the left or right) is already
very complex \cite{barbell}. Furthermore, phase spaces with different
parameters for the surrounding environment (e.g. friction coefficients) are
computationally intensive to be re-calculated every time when a condition is
changed. Note that if no bouncing occurs (e.g. a thick coin landing on a
soft-rice bed \cite{coin}), the phase space can be calculated; our work looks
into a more general case where bouncing is allowed.

Since the classical-mechanics approach is so complex as to be impracticable
and computationally intractable, we looked for a simpler method that could
utilize statistical principles. The Boltzmann distribution in statistical
mechanics attracted our attention as the principle requires only knowing an
ensemble of energy states and temperature to determine the distribution in
each state of particles. By constructing a partition function based on this
Boltzmann factor, $e^{-\varepsilon_i/k_BT})$ where $\varepsilon_i$ is the
energy in each state i, and $T$ is temperature, one can predict many
thermodynamic properties such as heat capacity of a solid and magnetic
susceptibility as a function of temperature \cite{Pathria,Goodstein}. The
concepts of statistical mechanics are also useful at large length scales
(e.g. star distribution \cite{stars}, with proposed entropic forces
\cite{entropic}) used for objects such as stars and black holes. In this
work, we are interested in applying the Boltzmann concept to a set of tossed
dice in 23 different dimensions (see results in Fig. 2). The redefined
temperature term, which is important for describing the outcome, is explained
in a later section (see eq. (13)).

\begin{figure}
\includegraphics [width=3.2in, clip]{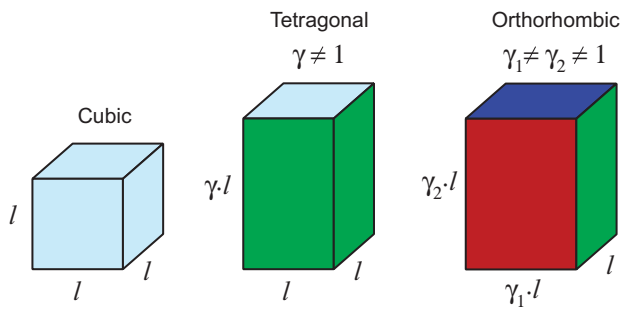}
\caption{\label{fig1} Geometry of regular dice (left), tetragonal dice (middle)
and orthorhombic dice (right). The parameters $\gamma$, $\gamma_1$ and $\gamma_2$
are the ratios between extended/contracted sides and the common length $l$.}
\end{figure}

For experimental measurements, the die shape and parameters are defined as
follows: Fig. 1 shows a cubic die with length of each side $l$. For the
tetragonal and orthorhombic dice, $\gamma$, $\gamma_1$ and $\gamma_2$ are
dimensionless ratios between extended/contracted sides and the common length
$l$, describing the dimensions of each die. All dice are custom-made from
solid aluminum. With $l$ of 0.9 cm, there are 11 sizes of tetragonal dice
with varied $\gamma$ as displayed along x-axis in Fig. 2(a) and 11 sizes of
orthorhombic dice with fixed $\gamma_1$ of 1.3 and varied $\gamma_2$ as
displayed along x-axis in Fig. 2(b). For each dimension, we toss the dice
around 2400 times on average by dropping them from a height of 27 cm above a
leveled and flat floor composed of thick and hard ceramic. Note that the
height from which the dice are dropped is many times larger than the
dimension of each dice. At this height, random outcomes are expected and the
regular die gives a 1/6 measured probability as expected due to symmetry
(please see supplementary materials for more explanation \cite{sup}). To
avoid biased measurements, the 33 persons, who tossed the dice, were not told
about the prediction results beforehand.

As shown in Fig. 2(a), we have measured the final configuration probabilities
of tossed tetragonal dice by varying the $\gamma$ factor. The outcome appears
to have a well-organized structure for the probability when plotted as a
function of the dimensionless length ratio $\gamma$. As the die is colored
with blue and green on 2 square faces and 4 rectangular faces respectively,
the blue region in Fig. 2(a) represents the probability of the blue face
being up (called $\gamma$ state) and vice versa for the green region; for
examples, a) if $\gamma$ is nearly zero (i.e. a thin square sheet), the
probability in observing the blue face up should be nearly 1 b) if $\gamma$
is large (i.e. a long stick with square cross-section), the probability of
finding the green face up should also be nearly 1 and c) if $\gamma$ is
exactly one (i.e. a cubic), the probability in landing with a blue(green)
face up should theoretically be around 1/3 and 2/3 respectively due to the
symmetry. For the orthorhombic dice, the probability outcome as a function of
$\gamma_2$ is shown in Fig. 2(b). The blue region represents the probability
of the blue face being up (the $\gamma_2$ state) and the same is applied for
the red and green regions.

\begin{figure}[t]
\includegraphics [width=3.4in, clip]{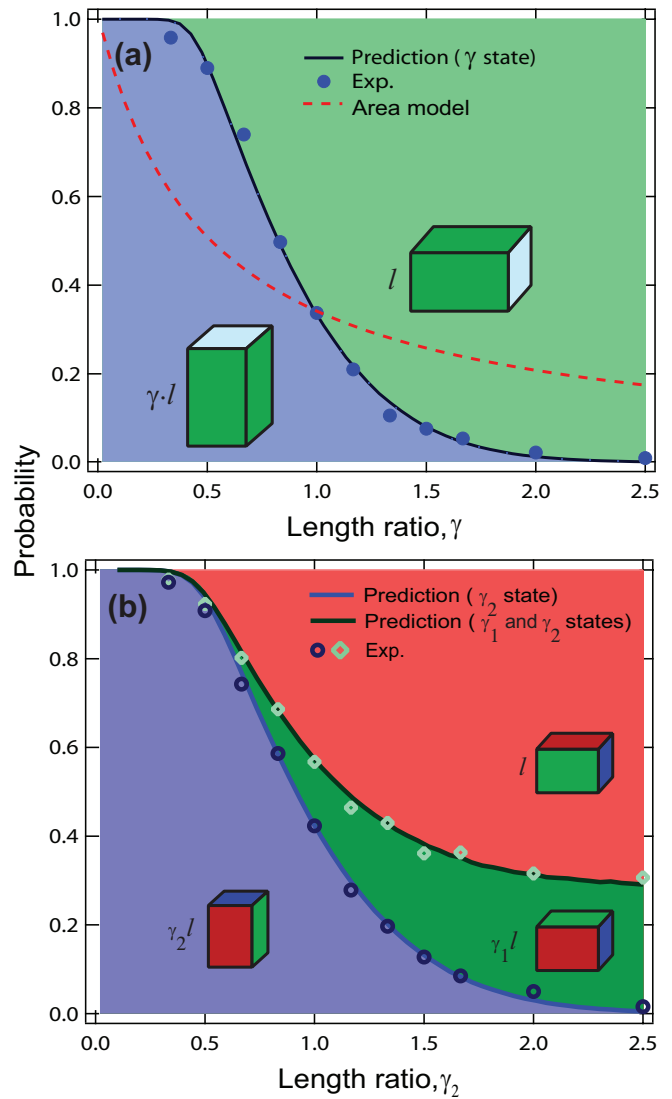}
\caption{\label{fig2} Probability distribution of landed dice
being in different orientations shown in the inset figures for (a) tetragonal dice
and (b) orthorhombic dice. The solid lines are model based predictions.}
\end{figure}

To model these experimental results, one could simply presume that the
probability outcome is directly proportional to the area of the face being
up. For example, in the case of a tetragonal die with dimension of
1$\times$1$\times$1.5 cm$^3$ (i.e. $\gamma$ = 1.5), the area of blue faces
will be 1+1 = 2 cm$^2$ while the total area of all faces will be
1+1+1.5+1.5+1.5+1.5 = 8 cm$^2$; hence this area model will predict that the
probability of getting blue faces should be 2/8 = 0.25 while the measured
probability is only around 0.076. In Fig. 2(a), this area model for the same
range of $\gamma$ is plotted. This area model is clearly erroneous except for
around $\gamma$ = 0 and 1.

In this work, a statistical-mechanics approach is chosen and the concept of a
Boltzmann distribution is applied. Through rederiving the Boltzmann
distribution function we can translate and define the relevant parameters,
including energy states and temperature, which are relevant to predicting and
understanding our experimental results.

The Boltzmann distribution of a system at equilibrium with a total number of
particles $N = \sum _i n_i$ and total energy $E = \sum _i \varepsilon_i n_i$
(for all energy states $\varepsilon_i$) can be derived by maximizing the
entropy $S = k \ln W$ \cite{BoltzmannDev}. The constraints are that $N = \sum
_i n_i = const$ and $E = \sum _i \varepsilon_i n_i = const$. Then one will
arrive that

\begin{equation}
n_i = e^{-\alpha} e^{-\beta\varepsilon_i}
\end{equation}

where $\alpha$ and $\beta$ are constants.

With $N = \sum _i n_i$ and definition of temperature T , we can arrive with

\begin{equation}
n_i =  \frac{N}{Z} e^{-\varepsilon_i/kT}
\end{equation}

where the partition $Z = \sum_i e^{-\varepsilon_i/kT}$. Details of the
derivation can be found in the supplementary information \cite{sup}.

In the continuous case, an average energy $\langle \varepsilon \rangle =
\langle \varepsilon(x_i, p_i) \rangle$ \cite{Pathria,Goodstein} can be
calculated as:

\begin{equation}
\langle \varepsilon \rangle =  \frac{\int\int \varepsilon \cdot e^{-\varepsilon(x_i,p_i)/kT}dx_idp_i}{\int\int
e^{-\varepsilon(x_i,p_i)/kT}dx_idp_i}.
\end{equation}

In this way, one can associate the temperature term with the average energy
instead of using the formal definition $T\equiv\partial E/\partial S$ which
may be harder to use in our case. For example, the average kinetic energy of
monatomic gas is:

\begin{equation}
\langle \varepsilon \rangle =  \frac{\int\int\int  \frac{p_x^2+p_y^2+p_z^2}{2m} e^{-\frac{p_x^2+p_y^2+p_z^2}{2m kT}}dp_xdp_ydp_z}{\int\int\int
e^{-\frac{p_x^2+p_y^2+p_z^2}{2m kT}}dp_xdp_ydp_z}=\frac{3}{2}kT
\end{equation}

where $\varepsilon = (p_x^2+p_y^2+p_z^2)/2m$; hence, the temperature term $kT
= 2 \langle \varepsilon \rangle/3$.

\begin{figure}[t]
\includegraphics [width=3.3in, clip]{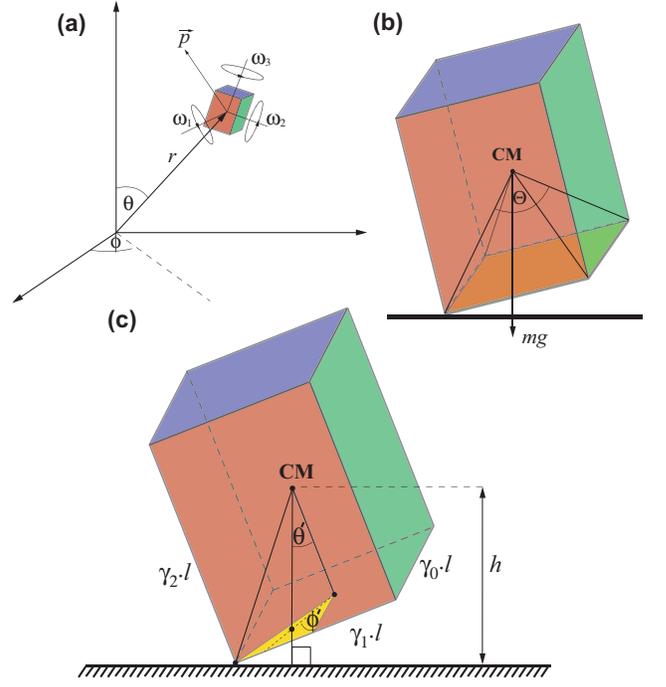}
\caption{\label{FIG3:diagram} Schematic diagrams of (a) a die with degrees of
freedom for calculating the total energy (eq. (5)), (b) the solid angle where
the weight vector points within and the final configuration will end up in
$\gamma_2$ state,  and (c) defined variables for integrations in eq. (8).}
\end{figure}

In our case, the total energy of a die, which includes a) translational
kinetic energy, b) rotational kinetic energy and c) potential energy (see
Fig. 3(a)), can be written as:

\begin{equation}
\varepsilon_d =\varepsilon(\omega_i, p_i, r, \theta, \phi) = \sum_{i=1}^{3} \frac{p_i^2}{2m} + \sum_{j=1}^{3} \frac{1}{2}I_j \omega_j^2 + mgr \cos{\theta}.
\end{equation}

To avoid confusion with the common temperature, we will replace the term $kT$
by $\kappa\tau$ which will be used as the redefined temperature for large
objects from now on. By using eq. (3), the average energy $\langle
\varepsilon_{d} \rangle$ as a function of $\kappa\tau$ can be calculated as:

\begin{equation}
\begin{split}
  \langle \varepsilon_{d} \rangle & =  \frac{\int\int\int  \varepsilon(\omega_i, p_i, r, \theta, \phi)\cdot e^{-\varepsilon(\omega_i, p_i, r, \theta, \phi)/\kappa\tau}d^3pd^3\omega d^3r}{\int\int\int e^{-\varepsilon(\omega_i, p_i, r, \theta, \phi)/\kappa\tau}d^3pd^3\omega d^3r}\\
  & = 6 \kappa\tau.
\end{split}
\end{equation}

The average energy can now be written as a function of $\kappa\tau$ (i.e. eq.
(6) whose detailed derivation is in Supplemental Material \cite{sup}). In
general, the $\kappa\tau$ term is arbitrary. However, when a specific
condition is given, the $\kappa\tau$ term can be calculated via a self
consistent equation. This is a crucial step. When the orthorhombic dice in
Fig. 2(b) were tossed, the total energy of the dice started from the high
value and then dissipated until coming to rest in either blue, green or red
configuration. In the case that the dice ended up being in blue region
($\gamma_2$ state), potential energy at some point in time (when kinetic
energy is zero) must be in the configuration that a die has one corner (or
more) touching the floor while the weight vector ($mg$) points down within
the solid angle $\Theta$ as shown in Fig. 3(b); then the die may move around
some more and eventually dissipates energy until it reaches the $\gamma_2$
state. These particular configurations give us the condition to calculate the
average energy as follow.

\begin{equation}
  \langle \varepsilon_{d} \rangle_{general} =  6 \kappa\tau = \langle \varepsilon_{d} \rangle_{constrained}.
\end{equation}

In our case, since the kinetic energy at that moment is zero and one corner touches the floor, we need to integrate only in angular space ($\theta$, $\phi$):

\begin{equation}
  \langle \varepsilon_{d} \rangle = 6 \kappa\tau =  \frac{\int  \varepsilon(\theta, \phi)\cdot e^{-\varepsilon(\theta, \phi)/\kappa\tau}d\Theta}{\int  e^{-\varepsilon(\theta, \phi)/\kappa\tau}d\Theta}.
\end{equation}

Instead of using the observer frame, the integral over the triangular solid
angle (yellow region) in Fig. 3(c) is done in the object frame
($\acute{\theta}$, $\acute{\phi}$):
\begin{equation}
\begin{split}
Ie(\gamma_0, \gamma_1, \gamma_2, \kappa\tau) & =\int  \varepsilon\cdot e^{-\varepsilon/\kappa\tau}d\Theta\\
& =  \int  mg\cdot h(\acute{\theta}, \acute{\phi})\cdot e^{-mgh/\kappa\tau} \sin(\acute{\theta})d\acute{\theta} d\acute{\phi}\\
Iz(\gamma_0, \gamma_1, \gamma_2, \kappa\tau) & =\int  e^{-\varepsilon/\kappa\tau}d\Theta\\
& = \int e^{-mgh/\kappa\tau}\sin(\acute{\theta})d\acute{\theta} d\acute{\phi}
\end{split}
\end{equation}
where
\begin{eqnarray}
\acute{\phi} & \in & \{0, \tan^{-1}{(\gamma_1/\gamma_0)} \},\\
\acute{\theta} & \in & \{0, \tan^{-1}{(\gamma_0/\gamma_2 \cos{\acute{\phi}})} \},   \\
h & = & \frac{l}{2}(\gamma_0 \sin{\acute{\theta}}\cos{\acute{\phi}}+\gamma_1 \sin{\acute{\theta}}\sin{\acute{\phi}}+\gamma_2 \cos{\acute{\theta}})
\end{eqnarray}
whose detailed derivations can be read from Ref. \cite{sup}.

To calculate the average energy, we need to integrate all the 6 faces of the
die. This can be calculated using eqs. (8)-(11) where the parameters
$\gamma_0$, $\gamma_1$ and $\gamma_2$ can be swapped for the integration on
each face (e.g. the yellow triangle in Fig. 3(c)) and for each configuration
of $\gamma_0$, $\gamma_1$ and $\gamma_2$, there are 8 equivalent pieces on
the die surface. By summing $Ie$ and $Iz$ for all configurations, covering
all the die surface, the eq. (7) can be written as:

\begin{equation}
\begin{split}
  \langle \varepsilon_{d}(\kappa\tau) \rangle & = \frac{\sum_{i\neq j \neq k} 8Ie(\gamma_i, \gamma_j, \gamma_k, \kappa\tau)}{\sum_{i\neq j \neq k} 8Iz(\gamma_i, \gamma_j, \gamma_k, \kappa\tau)} = 6 \kappa\tau
\end{split}
\end{equation}

where each index runs from 0 to 2.

\begin{figure}[t]
\includegraphics [width=3.3in, clip]{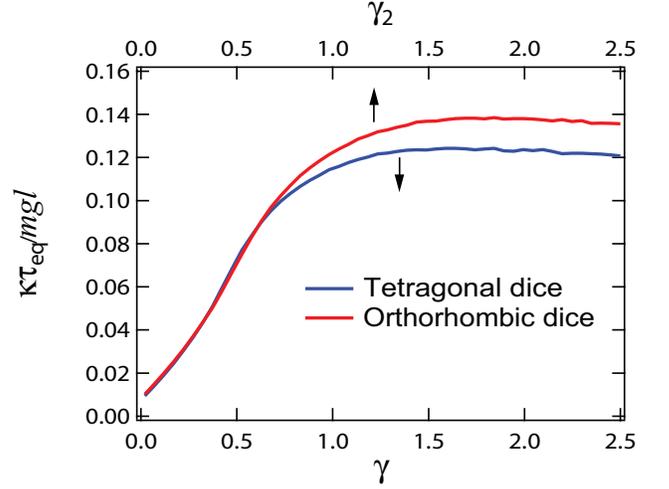}
\caption{\label{FIG4:temp} Temperature}
\end{figure}

From this eq. (12), the $\kappa\tau_{eq}$ for each $\gamma_2$ of the
orthorhombic dice can be calculated numerically as shown in Fig. 4. And by
plugging in these calculated temperature terms $\kappa\tau_{eq}$, the
probability in getting blue, green or red face of the orthorhombic dice is
given by:

\begin{equation}
\begin{split}
  p_{orth, blue} & = \frac{Iz(\gamma_0, \gamma_1, \gamma_2, \kappa\tau_{eq})+Iz(\gamma_1, \gamma_0, \gamma_2, \kappa\tau_{eq})}{Z_{orth}}\\
  p_{orth, green} & = \frac{Iz(\gamma_0, \gamma_2, \gamma_1, \kappa\tau_{eq})+Iz(\gamma_2, \gamma_0, \gamma_1, \kappa\tau_{eq})}{Z_{orth}}\\
  p_{orth, red} & = \frac{Iz(\gamma_1, \gamma_2, \gamma_0, \kappa\tau_{eq})+Iz(\gamma_2, \gamma_1, \gamma_0, \kappa\tau_{eq})}{Z_{orth}}\\
\end{split}
\end{equation}

where $Z_{orth}=\sum_{i\neq j \neq k} Iz(\gamma_i, \gamma_j, \gamma_k,
\kappa\tau_{eq})$, $\gamma_0$ = 1 and $\gamma_1$ = 1.3.

For the tetragonal dice, the calculations are the same as the case of the
orthorhombic dice except setting $\gamma_0 = \gamma_1 = 1$. The temperature
term for the tetragonal dice is also plotted in Fig. 4. It should be
emphasized that the normalized temperature term $\kappa\tau_{eq}/mgl$ only
depends on the shape (i.e. dimensionless $\gamma$ terms) but not the mass and
the length. Hence eq. (13) will give the same result of probability for the
object with the same shape regardless of its size. This is consistent with
the classical mechanics where the final configuration should not depend on
mass or length of the object (e.g. a freely-falling object always accelerates
at $g$).

Finally, the predictions from eq.(13) are calculated by using the
$\kappa\tau_{eq}/mgl$ shown in Fig. 4. The predictions are then plotted in
Fig. 2(a) and 2(b). The comparison with the experimental measurements shows
excellent agreement with coefficients of determination ($R^2$): 0.996 for
$\gamma$ state in Fig. 2(a), 0.997 and 0.995 for $\gamma_2$ and
$\gamma_1+\gamma_2$ state in Fig. 2(b) respectively.

With the statistical approarch providing for high accuracy predictions, it
could open the door to many applications. Without having to resort to
calculating ensembles of full trajectories, the Boltzmann formalism with
redefined temperature can be used as a simple and tractable method for
explaining and understanding the probability distribution of the final
configuration of moving macroscopic objects which eventually come to rest,
e.g. saving the cost of computation and experimental testing. This method may
also find application to the packing problems
\cite{packingDice,packingPRL,packingSci,packingMath}. By pouring objects into
a container (without shaking), the packing factor may be difficult to
calculate without knowing the probability of each object landing in a
particular configuration.

We would like to thank Ben Segal, Jan Zaanen and David Ando for the ideas
initiating this work and suggestions. This work was supported by Thailand
Research Fund (TRF) and Suranaree University of Technology (SUT) (Grant No.
BRG5880010) and Research Fund for DPST Graduate with First Placement (Grant
No. 021/2555). S.Siriroj acknowledges the Royal Golden Jubilee Ph.D. Program
(Grant no. PHD/0007/2555).


\begin{thebibliography}{99}

\bibitem{barbell} J. Nagler and P. Richter, Phys. Rev. E. 78, 036207 (2008).

\bibitem{coin} E.H. Yong and L. Mahadevan, Am. J. Phys. 79, 1195 (2011).

\bibitem{stars} J. N. Bahcall and R. A. Wolf. Astron. J. 209, 214-232 (1976)

\bibitem{entropic} E. Verlinde, J. High Energy Phys.  1104, 29 (2011).

\bibitem{Pathria} R. K. Pathria, "Statistical mechanics," Butterworth
    Heinemann, Oxford, UK, 1972.

\bibitem{Goodstein} D. L. Goodstein, âStates of Matter,â Prentice-Hall,
    Englewood Cliffs, NJ, 1975.

\bibitem{BoltzmannDev} Derivation of Boltzmann distribution is from the
    lecture note of Prof. Angel C de Dios; Also, H. A. Bent, "The Second
    Law," Oxford University Press, New York, 1965.

\bibitem{sup} See Supplemental Material at [URL] for probability as a
    function of dropping height and mathematical derivations.

\bibitem{packingPRL} A. Donev, F. H. Stillinger, P. M. Chaikin and S.
    Torquato, Phys. Rev. Lett. 92, 255506 (2004)

\bibitem{packingSci} A. Donev, I. Cisse, D. Sachs, E. A. Variano, F. H.
    Stillinger, R. Connelly, S. Torquato, P. M. Chaikin, Science. 303,
    990â993 (2004)

\bibitem{packingDice} A. Jaoshvili, A. Esakia, M. Porrati and P. M. Chaikin,
    Phys. Rev. Lett. 104, 185501 (2010)

\bibitem{packingMath} C. Zong, Expo. Math. 32, 297â364 (2014)

%


\end{thebibliography}
\end{document}